\documentclass[article,aps,floats,nofootinbib, floatfix,superscriptaddress]{revtex4}
\usepackage{graphicx} \usepackage{bm} \usepackage{epsfig}
\usepackage{pstricks}
\usepackage{amsmath}

\begin{document}

\title{The Private Higgs}

\author{Rafael A. Porto}
\affiliation{Department of Physics, University of California, Santa Barbara, CA 93106}

\author{A. Zee}
\affiliation{Department of Physics, University of California, Santa Barbara, CA 93106}
\affiliation{Kavli Institute for Theoretical Physics, University of California, Sana Barbara, CA 93106}

\begin{abstract}
We introduce Higgs democracy in the Yukawa sector by constructing a model with a private Higgs and a dark scalar for each fermion thus addressing the large hierarchy among fermion masses. The model has interesting implications for the LHC, while the Standard Model phenomenology is recovered at low energies. We discuss some phenomenological implications such as FCNC, new Higgses at the TeV scale and dark matter candidates. 
\end{abstract}

\maketitle

\section{INTRODUCTION}

Quark and lepton masses vary enormously. In the Standard Model (SM) of particle physics with a single Higgs, there is an enormous hierarchy among the Yukawa couplings $y_q$: for instance in the quark sector we have ${m_t \over m_u} = {y_t \over y_u} \sim 10^5$. Except for the top quark, $m_t \sim 170$ GeV, fermion masses are way below the Electroweak Symmetry Breaking (EWSB) scale $v_h \sim 246$ GeV \cite{djoudi}. We propose a simple way to address this Yukawa hierarchy in a democratic way, by introducing one Higgs per fermion, which we call a private Higgs (PH). In addition we also introduce a real scalar per fermion, the ``darkons",  blind to the SM quantum numbers.  
Here we introduce one particular realization of this general idea. It should be simple to construct variants of our specific model. The phenomenological implications are rich, given the plethora of new Higgses. The SM predictions are however recovered plus corrections due to an extended scalar sector. Many of our model's predictions are testable at the Large Hadron Collider (LHC). Furthermore, a dark sector provides missing energy channels and dark matter candidates. As in multi-Higgs model the issue of Flavor Changing Neutral Currents (FCNC) appears. We discuss possible solutions and constraints. 

In this paper we will not deal with the celebrated ``Hierarchy" puzzle in the scalar sector, namely why the electroweak scale is much smaller than the Planck scale. However, the strong indications that we live in a accelerating universe with an equally tuned cosmological constant may suggest that we are not looking at this issue in the proper way (it has been remarked, for instance, that the ratio between the cosmological constant and the electroweak scale is roughly of the same order of magnitude as that between the electroweak scale and the Planck scale). We will not adopt here the small ratio between the electroweak and Planck scale as a guidance\footnote{We cannot discard the possibility that it may be that whatever ``solves" the Hierarchy problem and stabilizes the Higgs mass could also apply to the PH model.}, but instead allow for significant extensions of the scalar sector (which has perhaps not been explored in depth so far) in order to address the other remaining puzzle in the SM, namely the observed hierarchy between fermion masses.

\section{THE PRIVATE HIGGS MODEL}

In this paper we focus on quarks, leptons are considered in \cite{ph2}. Our model has the 
same content as the SM but without the standard Higgs. Instead we introduce a PH field for each quark (and eventually  for each fermion), $\phi_q$ for $q=u,d,s,c,t,b$, transforming the same way under $SU(2) \times U(1)$ as the SM Higgs. As a variant of the model presented here, we could also include another Higgs for the gauge bosons $W,Z$. However, as we shall see, this would not be necessary and we could make do with the Higgs for the top quark.
To avoid cross talk between different quarks we introduce, similar to the Glashow-Weinberg model \cite{gwm}, a set of six separate discrete symmetries $K_q$ under each of which a set of $SU(2)\times U(1)$ real scalar fields $S_q$ also participates:

\begin{equation}
D_{\hat q}\to -D_{\hat q},  ~\phi_{\hat q} \to -\phi_{\hat q},~ S_{\hat q} \to -S_{\hat q}, 
\end{equation}
for ${\hat q}=(d,s,b)$, and

\begin{equation}
~ U_{\tilde q} \to - U_{\tilde q}, ~ \phi_{\tilde q} \to -\phi_{\tilde q},~ S_{\tilde q} \to -S_{\tilde q}, 
\end{equation}
for $\tilde q=(u,c,t)$. Here $U$ and $D$ denote right handed quark fields. Evidently, $S_{q}\phi_{q}$ is invariant under all the $K_q$'s. As we shall see these discrete symmetries are not enough to prevent tree level FCNC. We will discuss the phenomenological implications later on.\\

Our model is specified by the Lagrangian (here ${\cal L}_{SM}$ corresponds to the SM Lagrangian with the standard Higgs $H$ set to 0) 

\begin{widetext}
\begin{eqnarray}
{\cal L} &=& {\cal L}_{SM} + \sum_q \partial_\mu S_q \partial^\mu S_q -\frac{1}{2} M_{S_q}^2 S_q^2 - \frac{\lambda_s^q}{4} S_q^4 +
\left(   (D_\mu \phi_q)^\dagger D^\mu \phi_q - \frac{1}{2} M_{\phi_q}^2 \phi_q^\dagger\phi_q - \lambda_q (\phi_q^\dagger\phi_q)^2 + g_{sq} S_q^2 \phi_q^\dagger\phi_q
 \right) \nonumber\\
&+& \sum_{q \neq q'} \left(a^s_{qq'} S_q^2 S_{q'}^2+\gamma_{qq'} S_q S_{q'} \phi_q^\dagger\phi_{q'} +  \chi_{qq'} S_{q'}^2 \phi_q^\dagger\phi_q+ a_{qq'} \phi_q^\dagger \phi_{q'} \phi_q^\dagger \phi_{q'} + b_{qq'} \phi_q^\dagger \phi_q \phi_{q' }^\dagger \phi_{q'} + c_{qq'} \phi_q^\dagger \phi_{q'} \phi_{q' }^\dagger \phi_{q}   \right)\nonumber \\ &-&  \sum_{\hat qp}  Y^{p \hat q}_{D} {\bar Q}^{p}  \phi_{\hat q} D_{\hat q} - \sum_{\tilde q  p}Y^{p \tilde q }_{U} {\bar Q}^p \tilde\phi_{\tilde q} U_{\tilde q} + \mbox{h.c.}
\label{ls}
\end{eqnarray}
\end{widetext}
where  ${\tilde \phi}_q = i \sigma_2 \phi_q$. Also $p = (1,2,3)$ is a family index,  ${\hat q}= (d,s,b)$, $\tilde q = (u,c,t)$, and  $Y^{PH}_D, Y^{PH}_U$ are Yukawa matrices.\\

We could of course study this model systematically by a numerical approach, but it might be clearer to analyze the rather complicated Higgs and scalar sector by successive approximation. Thus,  we will explore some, but not all, of this model's parameter space, and provide a rough estimate of the model's virtues and phenomenological consequences. In principle there could be many regions of parameter space experimentally allowed in our model. 

To unclutter our analysis we will take the parameters $a_{qq'}, b_{qq'}, c_{qq'}$ to be negative and small and ignore them in what follows. We will drag along the term proportional to $a^s_{qq'}$ although we will subsequently assume it to be small. We take $M_{\phi_q}^2>0$ and induce EWSB  through the $g_{st},\chi_{qt},\gamma_{tq}$ couplings and the vacuum expectation value (vev) of the darkon fields $S_q$. We start with the PH for the top quark that we will identify with the SM Higgs.
Taking $\frac{1}{2} M_{\phi_t}^2 - g_{st} \langle S_t \rangle^2 -\sum_{q \neq t}  \chi_{qt} \langle S_q \rangle^2   < 0$, we force $\phi_t$ to develop a standard `negative mass squared' instability. The $SU(2)\times U(1)$ is thus spontaneously broken and the gauge bosons acquire mass. To give the $S_q$ fields a vev  we introduce an instability $M^2_{S_q} < 0$ such that the relevant pieces in the potential for $S_q,\phi_t$ become (notice that in principle $v^q_s$ is a ``bare" parameter)

\begin{equation}
\sum_q \left\{ {\frac{\lambda^q_s}{4}}\left(S_q^2- \frac{(v^q_s)^2}{2}\right)^2+\left[\frac{1}{2}M_{\phi_t}^2-\left(g_{st}\delta_{qt}+(1-\delta_{qt})\chi_{tq}\right) S_q^2\right]\phi_t^\dagger\phi_t \right\} + \sum_{q\neq q'} a^s_{qq'} S_q^2 S_{q'}^2  + \lambda_t (\phi_t^\dagger\phi_t)^2.
\end{equation}

We get, after minimization, 
\begin{eqnarray}
0 &=& \lambda_t \langle \phi^0_t \rangle^2 - \sum_q \frac{1}{2} \left(g_{st}\delta_{qt}+(1-\delta_{qt})\chi_{tq}\right) \langle S_q\rangle^2, \\
0&=&  \lambda^q_s \left( \langle S_q \rangle ^2 - \frac{(v^q_s)^2}{2} \right) - 2 \left(g_{st}\delta_{qt}+(1-\delta_{qt})\chi_{tq}\right)\langle \phi^0_t \rangle^2 + 2\sum_{q'\neq q} a_{qq'} \langle S_{q'}\rangle^2,
\end{eqnarray}
where we assumed $M_{\phi_t}^2 \ll g_{st} \langle S_t \rangle^2$.\\

For simplicity and illustrative purposes, we consider $g_{st} \sim \chi_{qt}$, $a_{qq'} \ll 1$, $\lambda_s^q \lambda_t \gg  \mbox{max} (g_{st}^2,\chi^2_{qt})$. Also, we assume that $\langle S_q \rangle \sim \langle S_t \rangle$ for all $q$'s,  $ v^q_s \sim v^t_s$. We could have equally introduce a hierarchy between the vevs of the $S_q$ fields. As we shall see that  also has important advantages. However, here we will assume the vevs are roughly equal and therefore we have 

\begin{eqnarray}
\langle \phi^0_t \rangle^2 &\sim& \frac{3}{\lambda_t}  g_{st} \langle S_t \rangle^2 \sim \frac{v_h^2}{2}\\
\langle S_q \rangle^2 &\sim& \frac{(v^q_s)^2}{2} \sim \frac{(v^t_s)^2}{2} ,
\end{eqnarray}
If, as suggested earlier, we identify $\phi_t$ with the SM field $H$, and $\phi_t^0$ with the Higgs particle $h$, we take $v_h \sim 246$ GeV. In this paper we will keep only leading order terms in $g_{st}$, so that $v_h \sim (\frac{3g_{st}}{\lambda_t})^{1/2} v^t_s$. It does not require a large hierarchy to enforce this condition. For instance, the mass of the ``SM Higgs" $(m_h)$  to leading order in $g_{st}$, is given by $m^2_h= 2 \lambda_t v_h^2 \sim 6 g_{st} (v^t_s)^2$. Therefore, if $v_h \sim 246$ GeV and we take $m_h \sim \left[\frac{v_h}{2},v_h\right]$, then we have $\lambda_t \sim \left[ \frac{1}{8},\frac{1}{2}\right]$. Therefore we take $g_{st} \sim \frac{\lambda_t}{3}$, and $\lambda_s \sim 1$. Notice that the latter is within the perturbative regime, which starts to break down around $\lambda_s \sim 4\pi$.  Under these assumptions ${ 3 g^2_{st} \over \lambda_s\lambda_t  }\sim \frac{g_{st}}{\lambda_s} \sim [\frac{1}{24},\frac{1}{6}]$ and our approximations are roughly at the 10 percent level depending on the value for $m_h$. Note we also have $v^t_s \sim v_h$. All these values could be changed of course depending on the choice of parameters and the experimental outcomes.\\
 
Here we have written the vevs as real quantities for simplicity. With this many scalar fields around, we have plenty of opportunity to allow for CP violation beyond phases in the CKM matrix. We could easily allow for relative phases in the vevs.  
 
We next look at the term $\gamma_{qq'} S_q S_{q'} \phi_q^\dagger\phi_{q'}$ in the Lagrangian. If we write $S_q = \frac{1}{\sqrt{2}}(v^q_s+ \sigma_q)$ and $\phi_t^0 =\frac{1}{\sqrt{2}}(v_h+h)$, after the fields acquire their respective vacuum expectation values, this term will induce pieces which are linear in $\sigma_q, h$  and the real part of the field associated with $\phi_q$, and quadratic mixing between them. Again, instead of diagonalizing a large matrix we will content ourselves with successive approximations in order to see more clearly what is going on. It will turn out that, in the region of parameter space we are looking at, most of these terms lead to small corrections. The most important term is the term linear in $\phi_q^0$ to which we now turn.

We have a choice regarding how the $\phi_{q \neq t}$ fields acquire vevs. We could have the usual `negative mass squared' instability, but since we already have a term linear in $\phi_q^0$ available, we opt to impose  $M_{\phi_q} >  \sqrt{g_{sq}} v^q_s$, for ${q \neq t}$. Thus, while it would be interesting to explore this model in other regions of parameter space, in this paper we will drive the spontaneous symmetry breaking of these other Higgs fields via the ${\gamma_{qt}}$ couplings. When $S_q$ and $\phi_t^0$ pick up vevs, the pieces 
\begin{equation}
\frac{1}{2}M_{\phi_q}^2 \phi^\dagger_q\phi_q - \frac{\gamma_{qt}}{\sqrt{2}} \frac{v^t_s v^q_s v_h}{2} \phi_q^0\label{vev0}
\end{equation}
in the potential will generate a vev for the neutral component of each of our private non-top Higgses ($q \neq t$)
\begin{equation}
\label{vev0q}
\langle \phi^0_q \rangle = \gamma_{qt} \frac{v_h}{\sqrt{2}} \frac{v^t_s v^q_s}{2M_{\phi_q}^2} \sim \gamma_{qt} \frac{v_h}{\sqrt{2}} \frac{(v^t_s)^2}{2M_{\phi_q}^2}, 
\end{equation} 
where we used $v^q_s \sim v^t_s$.  

The expression in (\ref{vev0q}) allows us to avoid unnaturally small Yukawa couplings and hence realize the motivating idea behind the PH scheme. For each non-top quark we now have two parameters, $\gamma_{qt}$ and $M_{\phi_q}^2$ to play with. By having $\gamma_{qt}$ small and $M_{\phi_q}^2$ large (note that this is consistent with the condition we imposed above) we could make $\langle \phi^0_{q \neq t} \rangle$ small and hence all the Yukawa couplings to be of similar order. Notice that in general the similarity of the vevs in the $S_q$ sector can be relaxed, thus giving us even more freedom. For instance by considering $v^q_s \ll v^t_s$ we could in principle also generate small values of $\langle \phi^0_{q\neq t} \rangle$. Conversely we may also go in the opposite direction and take $v^q_s \gg v^t_s$ which would imply heavier PHs. 

\section{Induced quark mass ratios}

To quantify the benefits of our PH scenario we can define
\begin{equation}
\mbox{tan}\beta_q \equiv \frac{\langle \phi^0_t \rangle}{\langle \phi^0_q\rangle} = \left( \frac{2M_{\phi_q}^2}{\gamma_{qt} (v^t_s)^2}\right) \label{tanb}.
\end{equation}

The Yukawa couplings in the PH model can thus be compared with the SM values
\begin{equation}
y^{PH}_{q \neq t} = \mbox{tan}\beta_q ~ y^{SM}_{q\neq t}, ~~ y^{PH}_t = y^{SM}_t  \sim 1.
  \label{yph}
\end{equation}
Let $\tau_q  \sim \frac{m_t}{m_q}$, denote the desired value of $\mbox{tan}\beta_q=\frac{y^{PH}_q}{ y^{SM}_q}$, so that $\tau_q \gg 1$. Then we obtain
\begin{equation}
\label{mph}
 M_{\phi_q} \sim  \sqrt{\frac{\tau_q}{\xi_{sqt}}} v_h \sim \sqrt{\gamma_{qt} \tau_q} v_h,
\end{equation}
where we introduced $\xi_{sqt} \equiv \frac{3g_{st}}{\lambda_t \gamma_{qt}}$, and in the last step we used $v^t_s \sim  v_h$.\\

Our model has the interesting feature that the heavier the quark, the smaller $\tau_q$ and hence the lighter its associated PH particle. 
Consider for example the up quark, for which ${m_t \over m_u} \sim 10^{5}$. Taking   $\xi_{sut} \sim 1$ we obtain
$M_{\phi_u} \sim 10^{5/2}   v_h \sim 10^3  \mbox{TeV}$,
most likely outside LHC range. In contrast, with ${m_t \over m_b} \sim 40$, if $\xi_{sbt} \sim 1$, the PH associated with the $b$ quark appears at a mass scale which naturally falls in the few TeV range.\\

There are important phenomenological consequences from all these interactions, for instance in the mass basis the quarks will talk to each one of our PHs to a different degree. The main phenomenological applications will come from the PH associated with the $b$ quark. The other Higgses will be heavy enough to `decouple' in the `low energy' regime we will explore at LHC.
In what follows we will explore some basic phenomenological implications and constraints of our model, in particular the appealing feature of providing a dark matter candidate. We will discuss these in more detail in future work.

\section{PHENOMENOLOGY}

\subsection{Gauge Bosons}

Let us begin with the gauge bosons. At tree level, the $W$ boson acquires mass with 
\begin{equation}
m^2_W = \frac{1}{2} g_{ew}^2 v_h^2 \left( 1 + \sum_{q\neq t} \left(\frac{1}{\tau_q}\right)^2 \right),
\end{equation}
and similarly for the $Z$ boson. Note that the parameter $\rho = \frac{M_W}{M_Z \mbox{cos} \theta_W}$ stays equal to unity since our PH have the same quantum numbers under $SU(2) \times U(1)$ as the standard Higgs, and $S_q$ are $SU(2) \times U(1)$ singlets.

\subsection{Flavor Changing Neutral Currents}

Multi-Higgs models have been considered previously in the literature \cite{yuval}. In general they present the problem of tree level FCNC. The existence of tree level FCNC would require new physics at a very large scale, putting strong bounds on the parameters, or the mass scale of new physics, in any theory beyond the SM. With ${\cal O}(1)$ parameters we would need a scale of new physics as large as $10^3-10^4$ TeV \cite{ben}.

Besides supersymmetry, a very popular way out of this problem is the two Higgs Doublet Model (2HDM)\cite{thdm}. In a general framework, using the Glashow-Weinberg discrete symmetry (as in type II 2HDM) one can couple each quark of a given charge to one, and only one of the Higgs doublets and hence avoid tree level FCNC \cite{gwm}. Another recently explored possibility goes by the name of Minimal Flavor Violation (MFV) \cite{mfv,jure}. In a MFV scenario the physics beyond the SM is assumed to be invariant under the transformation
$
Q_L \to V_L Q_L, ~ U_R \to V_U U_R, ~ D_R \to V_D D_R,~
Y^D \to V_L Y^D V^\dagger_U,  ~ Y^U \to V_L Y^U V_D^\dagger 
$
where $Y^D$ and $Y^U$ are the Yukawa matrices. The SM is formally invariant under these transformations. MFV implies that FCNC, as in the SM, are naturally suppressed by small CKM mixing angles. A multi-Higgs theory with MFV has been recently studied in \cite{annesh,wise}.\\ 

In this section we will explore the issue of FCNC in the PH model and the possible venues towards a {\it natural} solution (without fine tuned parameters we wanted to avoid in the first place) in agreement with experiments. As we shall see, we will also attempt to connect the smallness of the CKM mixing elements with the suppression of FCNC, although it is clear that our model does not respect the MFV symmetry.

In our PH model the most general interaction between PHs and quarks takes the form (we suppressed the PH label),
\begin{equation}
\label{generalQ}
{\cal L}_{quarks} = -  \sum_{\hat qp}  Y^{p \hat q}_{D} {\bar Q}^{p}  \phi_{\hat q} D_{\hat q} - \sum_{\tilde q  p}Y^{p \tilde q }_{U} {\bar Q}^p \tilde\phi_{\tilde q} U_{\tilde q} + \mbox{h.c.}
\end{equation}

In accordance with the philosophy behind our model, after each one of the PH gets a vev, we would like the left handed mass eigenstates, with properly normalized kinetic pieces, to be approximately given by $({\bar d}^{\hat q}_{L})_{\rm ph} \equiv \sum_p Y^{p\hat q}_{D} {\bar d}^p_L$ and $({\bar u}^{\tilde q}_{L})_{\rm ph} \equiv \sum_{p} Y^{p\tilde q}_{U} {\bar u}^p_L$. However, that would require the two Yukawa matrices, $Y_D$ and $Y_U$, to be proportional to two approximately unitary matrices. That is not necessarily the case in the the general expression in (\ref{generalQ}). In principle we would need to diagonalize the mass matrix, for instance for $D$-type quarks, $M_D^{p\hat q} = Y^{p \hat q}_D \langle \phi_{\hat q}\rangle$ (no sum on $\hat q$). In the mass eigenstates basis the PHs will induce tree level FCNC, and furthermore it is not even guaranteed the natural hierarchy of masses will be preserved. 

There are two possible routes to avoid this problem. As we sketched above, either we rely on phenomenology, e.g. experimental constraints, and some degree of fine tuning, or the existence of a symmetry. There is an exhaustive history behind flavor symmetries \cite{zeewil} in general. However, flavor symmetries tend to be either too restrictive, as is the case with continuous non-abelian groups, or too loose, as with abelian or $Z_N$ type groups. A more appealing choice seems to be discrete non-abelian groups \cite{ma}, which have attracted some attention in recent years due to the discovery of neutrino masses \cite{a4}.  Here we will adopt a more phenomenological approach, inspired by the possible existence of an underlying flavor symmetry and the MFV spirit.\\

Recall that in the SM the CKM matrix elements are given by
\begin{equation}
V_{CKM} = V^D_L (V^U_L)^\dagger,
\end{equation}
where $V^U_L$ and $V^D_L$ are the unitary transformations to the left handed mass eigenstates basis.
If our $Y_U,Y_D$ matrices were unitary, that would immediately allow us to avoid any cross talking and the dreaded FCNC, however, at the same time that would lead to a completely general CKM structure, even more in the PH model where $Y_{D(U)} \sim {\cal O}(1)$. On the other hand, a natural way to obtain small CKM mixing angles is to consider $V^D \sim V^U$, in other words, the U-type and D-type quarks rotate in a similar fashion and the small mixing angles originate in some deviation from this leading order picture, perhaps due to the different mass ratios. This line of thought suggests the idea that CKM matrix elements and mass ratios may be connected. On the other hand, it is possible that mixing angles and quark masses are manifestations of different phenomena. We believe the PH scenario provides an open playground to study these issues. For instance, a possibility would be to construct a PH model with the $q=1\ldots 6$ Higgses, $\phi_q$, transforming under certain representation of a flavor group $G_F$ such that the symmetry will force the Yukawa interactions to avoid flavor cross talking. This symmetry is an approximate symmetry  in the low energy effective theory which would explain the smallness of the CKM mixing angles. Let the $G_F$ symmetry be broken at some scale $M_F$. The smallness of ${M_{PH} \over M_F}$, with $M_{PH}$ the scale of the PH model, may account for the smallness of flavor violation and the CKM elements. Based on these set of ideas we assume the existence of such $G_F$ symmetry which is broken leaving as a remnant the discrete subgroup generated by the $K_q$'s,  such that the Yukawa matrices in our theory could be assumed to take the form
\begin{equation}
\label{ydij}
 Y^{p\tilde q}_{U} = \lambda_U \delta^{p\tilde q} + \epsilon^{p\tilde q}_U, 
\end{equation}
and similarly for D-type quarks. In these expressions $\lambda_{D(U)}$ is an ${\cal O}(1)$ overall proportional constant and $||\epsilon^{pq}_{D(U)}|| \ll 1$. Notice the interesting effect that the small breaking of the $G_F$ symmetry in the low energy theory would also guarantee the smallness of the $\gamma_{qq'}$ couplings which generate the vevs of the PHs other than $\phi_t$ (although we could take the scalar fields $S_q$ also in some representation of $G_F$ such that the term proportional to $\gamma_{qq'}$ is a singlet under $G_F$; or equivalently consider only the quarks charged under $G_F$.). Note as well that in the SM the assumption in (\ref{ydij}) would imply totally degenerate quark masses. However, in the PH setting the mass ratio will be due to the different vevs of the PHs, up to an overall constant. The mass matrix for U-type quarks becomes (no sum over $\tilde q$)
\begin{equation}
 M^{p\tilde q}_U = \lambda_U \delta^{p\tilde q} \langle \phi_{\tilde q} \rangle + \epsilon^{p\tilde q}_U\langle \phi_{\tilde q} \rangle,
\end{equation}
and equivalently for D-type quarks. Within this framework we can naturally accommodate the quark mass hierarchy dynamically into the vevs of the PHs. Schematically the mass matrix takes the approximated form 
\[  M_U = \left( \begin{array}{ccc}
\lambda_U \langle \phi_u^0\rangle & \sim 0 & \epsilon \frac{v_h}{\sqrt{2}} \nonumber \\
\sim 0 & \lambda_U \langle \phi_c^0\rangle  & \epsilon \frac{v_h}{\sqrt{2}} \\
\sim 0 & \sim 0 & \lambda_U \frac{v_h}{\sqrt{2}} \nonumber \end{array} \right)\] 
with $||\epsilon^{p \tilde q}|| \sim \epsilon \ll 1$, and we neglected terms proportional to $ \epsilon \langle\phi^0_{u(c)}\rangle$ (recall $\langle \phi^0_q \rangle \sim \langle \phi^0_t \rangle/\tau_q$), but we keep the contribution from the top proportional to $\epsilon v_h$. Nevertheless, the mass eigenvalues are consistently, and naturally, given by the vevs of the PHs as we expect, and the transformation of the mass eigenstates will be approximately given by a similar form as the Yukawa matrix 
\begin{equation}
V^{D(U)}_L  = \mbox{Id} + \tilde \epsilon_{D(U)},
\end{equation}
with $||\tilde \epsilon_{D(U)}|| \sim \epsilon \ll 1$ describing a small deviation generated by $ \epsilon_{D(U)}$,
and the CKM by the mismatch between $V^U_L$ and $V^D_L$, which will be naturally small for small values of $\tilde \epsilon_{D(U)}$. A similar transformation applies for the right handed quarks. 
We intend to explore the internal hierarchy within the CKM matrix itself in a future work.\\ 

We now come to the FCNC constraints. Due to suppression effects for cross talking between different PHs, the most stringent bounds will come from neutral $B$-$\bar B$ and $K$-$\bar K$ mixing. For the $K$-$\bar K$ system the PHs involved are $\phi^0_d$ and $\phi^0_s$. For $B$ mixing we have the a priori lighter PH $\phi_b^0$ which will dominate the amplitude. The new physics effect would be thus naively suppressed only by $\Lambda \sim M_{\phi^0_b}$, which would imply $M_{\phi^0_b} \sim 10^3$ TeV. However, from the above discussion, there is an extra factor of $\epsilon^2$ in the amplitude. Notice that the experimental constraints translates into $\tilde\epsilon^{bd}_{D} \sim 0.05$ for the PH model to accommodate the bounds. This value is within the order of magnitude we would expect the elements in $\tilde \epsilon_{D}^{ij}$ to be at. Similar considerations apply for the Kaon system, for which the numerical bounds are very stringent. But now we also have a heavier Higgs particle $\phi^0_s$ involved, perhaps of the order
 $M^2_{\phi_s} \sim 40 M^2_{\phi_b}$. Keep in mind that is also possible to crank up the mass of the PHs by adjusting the vev for the $S_q$ fields in (\ref{vev0}).
 
 \subsection{Flavor Changing Charged Currents}
 
After EWSB we will also have charged Higgses $\phi^\pm_q$ floating around, and therefore flavor changing charged currents do appear. However, these are again suppressed by $\frac{1}{M_{\phi_q}^2}$.  From experimental constraints \cite{aleph,opal} one has \begin{equation}\frac{\mbox{tan}\beta_q}{M_{\phi^+_q}} \le 0.4 - 0.5 ~ \mbox{GeV}^{-1},\end{equation} and for the case of the $b$ quark it gives $M_{\phi^+_b} \ge 60 ~\mbox{GeV}$, which is naturally obeyed in our model. Other experiments put the lower bound around $300$ GeV \cite{circa}, which is also way below the scale of $M_{\phi^+_b}$. Notice some of these bounds come from leptonic decays, which are in addition suppressed in our model due to the small mixing between different PHs \cite{ph2}.\\ 

\subsection{Unitarity}

Another important constraint comes from unitarity bounds. Unitarity constraints for the Higgs particle have a long story, see \cite{djoudi} for references. Basically, partial wave unitarity constraints the mass of the SM Higgs, which unitarizes the theory for $WW$ scattering. With $A(WW \to WW) \sim G_F m_h^2$, unitarity implies $m_h \le$ 1 TeV. In our theory, $WW$ scattering with $h$ exchange will work out the same way. For the exchange of other PH we will have  $\frac{1}{\tau^2_q}$ suppression from twice the coupling $\phi_q WW$, and unitarity in $WW$ scattering is clearly obeyed. $WW$ scattering is not the only possible channel where bounds could apply. In the case of 2HDMs it has been shown the mass of the partner Higgs is severely constrained, again to be in the TeV scale, by different processes \cite{unit}. It is possible to show however that these bounds do not apply to our model, since the PH masses (other than $h$) are not generated via EWSB. The bounds are effectively constraints on the couplings in the potential, such as $a_{qq'},b_{qq'},c_{qq'}$, which can be always accommodated to satisfy unitarity without major effects.

\subsection{Mixing: Recovering the SM and LHC physics}

Let us write $\phi_{q\neq t} ^0 = \langle \phi_{q\neq t}^0 \rangle  + H_q + i A_q$. As mentioned earlier, the $\gamma_{qq'}$  term induces linear and quadratic terms between $h, H_q$ and $\sigma_q$. Let us discuss first the linear terms. We have already exploited the term linear in $\phi_q$ in (\ref{vev0}). The term linear in $h$ leads to a fractional shift in $v_h$ given by ${{\delta v_h}/v_h} \sim ({\gamma_{qt}/{\tau_q}})({{v^s_t}/{v_h}})$ (recall $v^s_q \sim v^s_t$) which with our choice of parameters is small. Similarly, the term linear in $S_q$ leads to a small shift in $v^q_s$ provided that  $({\gamma_{qt}/{ \tau_q}})\ll {\lambda^q_s}({v^s_qv^s_t}/{v_h^2})$. The condition that the mixing between $h$ and $\sigma_q$ be small compared to the mixing we already have  ( see (\ref{mixhs}) below) gives ${\gamma_{qt}/ \tau_q} \ll g_{st}$ which is readily satisfied for the parameter choices we have made. Similarly, the condition that the mixing between $\sigma_q$ and $H_q$ be small gives ${v_h/\tau_q}\ll v^q_s$, also readily satisfied. 

The mixing between $h$ and $H_q$, though small, leads to interesting phenomenological consequences because this induces a coupling between the non-top quarks and $h$. In the $H_q,h$ potential we have (for compactness we write $M_{H_q} \equiv M_{\phi^0_q}$)
\begin{equation}
\frac{1}{2} m_h^2 h^2 + \frac{1}{2} M_{H_q}^2 H_q^2 - \frac{\gamma_{tq}}{2} v^s_tv^s_q \frac{h}{\sqrt{2}} H_q.
\end{equation}
After diagonalization we thus get a $q \bar q h$ term with coefficient $\sim \frac{y^{PH}_q}{\sqrt{2}\tau_q} $, which implies that effectively we can mimic the SM interaction between $h$ and all the quarks. Therefore, the PH model reproduces the SM as far as the traditional Higgs is concerned, plus corrections due to the existence of the dark sector\footnote{More mixing arises from the other couplings in the Lagrangian, as well as mass shifts, for instance from the $a_{qq'},b_{qq'},c_{qq'}$ couplings, although we ignored these since they are doubly suppressed.}. 
Before delving into the latter let us explore some basic elements relevant for LHC physics, where we can concentrate in the $t,b$ quarks since lighter ones will have heavier PHs which will decouple at LHC energies. Let us focus on $h,H_b$. In the SM, the process $ b\bar b \to h$ is Yukawa suppressed. For hadron machines  we also have suppression in the parton distribution functions (PDFs). 
In our model, the most natural candidate to produce $H_q$ is via gluon fusion. The contribution to the partonic cross section follows the same steps as in the SM \cite{djoudi} with $m_h \to M_{H_b}$ and $y^{PH}_b \sim {\cal O}(1)$. To get the cross section we still have to integrate with the PDFs. Unfortunately we will also face luminosity suppression since $m_{H_b} \gg m_h$. Similar considerations apply for the CP-odd scalar $A_b$.

\section{the dark sector}

A very attractive aspect of our theory is the existence of a dark sector. Having a scalar particle responsible for dark matter was first proposed over twenty years in \cite{zee1,zee2} and later also in \cite{burg, PattWilczek, saopaulo, xgh}. In our model we have many of them, $\sigma_q$, which we called  ``darkons" following \cite{xgh}, and could naturally be the candidates for dark matter. We now have a blind set of scalars $\sigma_q$ and new interactions. This fields $\sigma_q$ couple to the Higgs sector in many ways. The Lagrangian in the $h,\sigma_q$ sector is given by
\begin{eqnarray}
 {\cal L}_{\sigma_q h} &=& \left(\sum_q \frac{1}{2}\partial_\mu \sigma_q \partial^\mu \sigma_q - \frac{1}{4} \lambda^q_s (v^q_s)^2 \sigma_q^2 - \frac{\lambda^q_s v^q_s}{4} \sigma_q^3 - \frac{\lambda^q_s}{16} \sigma_q^4\right) + \frac{1}{2}\partial_\mu h \partial^\mu h  - \lambda_t v_h^2 h^2  - \lambda_t v_h h^3 - \frac{\lambda_t}{4} h^4 \nonumber \\ &+& \frac{g_{st}}{4}[4v^t_s v_h \sigma_t h+ \left(\sigma_t^2 h^2 + 2 v_h\sigma_t^2h +2 v^t_s \sigma_t h^2\right)]  + \sum_{q\neq t} \frac{\chi_{qt}}{4}[4v^q_s v_h \sigma_q h+ \left(\sigma_q^2 h^2 + 2 v_h\sigma_q^2h +2 v^q_s \sigma_q h^2\right)] \nonumber \\ &+&   \frac{1}{2} \sum_{q' \neq q} a^s_{qq'} (v^q_s+\sigma_q)^2(v_s^{q'}+\sigma_{q'})^2.
 \label{mixhs}
\end{eqnarray}

The interaction between $h$ and $\sigma_q$ given by the last three terms has very interesting phenomenology. Notice that we could indeed totally ignore the mixing term between $h$ and $\sigma_t$ (and similarly for other $\sigma_q$'s) if $8g_{st}^3 \ll (2g_{st}-\lambda^t_s)^2 \lambda_t$ which is amply satisfied for the illustrative values chosen above.  The mass of the scalar bosons $\sigma_q$ is then given by 
\begin{equation}
m_{\sigma_q}^2 \sim \frac{\lambda^q_s}{2}  (v^q_s)^2  \to
m_{\sigma_q} \sim \frac{v_h}{\sqrt{2}} \sim m_t \sim 170~\mbox{GeV}~~~~ (\lambda_s^q \sim 1,~ a_{qq'}^s \ll 1).
\end{equation}
 
 Through its mixing with the PHs, the scalar sector couples  to quarks. Therefore, in general, depending on $m_{\sigma_q}$ we could have some amount of missing energy at the LHC. Notice that under our assumptions, $m_{\sigma_q} \sim m_t$, and thus $\sigma_q$ cannot decay into $hh$ since our current bounds on the Higgs mass set $m_h > 114$ GeV \cite{djoudi}.  That will give us a semi-stable 
 $\sigma_q$ \cite{chris}. On the other hand, a stable dark matter candidate may be provided by the darkon partner of the electron (see \cite{ph2} for details). 
 One important constraint is given by the Spergel-Steinhardt  bound \cite{ssbound}. As shown in \cite{zee2} this bound may be difficult to implement in a perturbative framework ($\lambda_s < 4\pi$) with a single dark matter scalar fields. But with the vast number of scalar particles and couplings here, we can readily accommodate existing experimental constraints.
 
 \section{Conclusions}

Here we constructed a private Higgs model which is natural in the Yukawa sector. It seems to agree with EW precision tests, at least in the region of parameter space we have looked at, and has some important phenomenology to be explored at LHC, like heavier Higgses for light quarks and a dark sector. The latter provides a candidate for dark matter. Also EWSB is driven by a discrete symmetry breaking. The most pressing issue is understanding the mechanism to suppress FCNC in a natural way. We believe that the PH model allows us to settle the question of flavor symmetries in a more natural fashion and could in principle elucidate a connection between the quark mass ratios and the smallness of the CKM matrix elements. We have extended our model to include leptons in \cite{ph2}. Most of the $\phi_l$ fields will be out of LHC's reach, except perhaps for the $\phi_\tau$. We will analyze these and other features in future work.\\

{\it Acknowledgments--} We thank one of the referees for very helpful comments.
AZ is supported in part by NSF under Grant No. 04-56556, and RAP by the Foundational Questions Institute (fqxi.org) under grant RPFI-06-18 and the University of California.

\end{document}